\documentclass[conference]{IEEEtran}
\IEEEoverridecommandlockouts
\usepackage[utf8]{inputenc} 
\usepackage[T1]{fontenc}
\usepackage{cite}
\usepackage{amsmath,amssymb,amsfonts}
\usepackage{algorithmic}
\usepackage{graphicx}
\usepackage{textcomp}
\usepackage{xcolor}
\usepackage{fancyhdr}
\usepackage[most]{tcolorbox}
\usepackage{booktabs}
\usepackage{multirow}
\usepackage{makecell}
\usepackage{adjustbox}
\usepackage{pifont}
\newcommand{\xmark}{\ding{55}}%
\usepackage{hyperref}
\newcommand{\codeicon}[1]{\href{#1}{\includegraphics[height=1em]{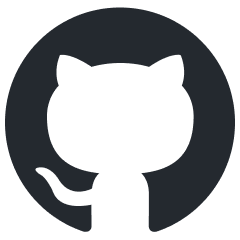}}}

\def\BibTeX{{\rm B\kern-.05em{\sc i\kern-.025em b}\kern-.08em
    T\kern-.1667em\lower.7ex\hbox{E}\kern-.125emX}}

\usepackage{fancyhdr}
\fancypagestyle{plain}{%
  \fancyhf{} 
}

\begin{document}

\title{SoK: Security and Privacy of AI Agents for Blockchain\\
\large \emph{This work has been accepted to the 7th International Conference on Blockchain Computing and Applications (BCCA 2025)}}

\author{
    \IEEEauthorblockN{Nicolò Romandini\IEEEauthorrefmark{1}, Carlo Mazzocca\IEEEauthorrefmark{2}, Kai Otsuki\IEEEauthorrefmark{3}, Rebecca Montanari\IEEEauthorrefmark{1}}
    \IEEEauthorblockA{\IEEEauthorrefmark{1}University of Bologna, Bologna, Italy}
    \IEEEauthorblockA{\IEEEauthorrefmark{2}University of Salerno, Fisciano, Italy}
    \IEEEauthorblockA{\IEEEauthorrefmark{3}NTT Digital, Inc. and NTT DOCOMO, Inc., Tokyo, Japan}
    \IEEEauthorblockA{Email: \{nicolo.romandini, rebecca.montanari\}@unibo.it, cmazzocca@unisa.it, kai.ootsuki.su@nttdocomo.com}
}

\maketitle

\begin{abstract}

Blockchain and smart contracts have garnered significant interest in recent years as the foundation of a decentralized, trustless digital ecosystem, thereby eliminating the need for traditional centralized authorities. Despite their central role in powering Web3, their complexity still presents significant barriers for non‑expert users. To bridge this gap, Artificial Intelligence (AI)‑based agents have emerged as valuable tools for interacting with blockchain environments, supporting a range of tasks, from analyzing on‑chain data and optimizing transaction strategies to detecting vulnerabilities within smart contracts. While interest in applying AI to blockchain is growing, the literature still lacks a comprehensive survey that focuses specifically on the intersection with AI agents. Most of the related work only provides general considerations, without focusing on any specific domain. This paper addresses this gap by presenting the first Systematization of Knowledge dedicated to AI‑driven systems for blockchain, with a special focus on their security and privacy dimensions, shedding light on their applications, limitations, and future research directions.
\end{abstract}


\section{Introduction}
The advent of blockchain and smart contracts has played a key role in shaping the Web3 ecosystem \cite{wu2023etal}, enabling a range of innovative services, ranging from decentralized identity management \cite{mazzocca2025survey} to decentralized finance (DeFi) \cite{CHEN2020e00151}. However, despite their significant potential, several limitations still prevent these technologies from reaching widespread adoption \cite{8343163}. Technical barriers often make it challenging for non-technical users to effectively utilize wallets or understand the basic concepts of smart contracts. Moreover, the abundance of available information can overwhelm individuals, making it difficult for them to identify the solutions best suited to their needs. At the same time, security and privacy concerns, such as phishing attacks, key mismanagement, and data leakage, remain a critical obstacle to gaining user trust and widespread acceptance.
\begin{figure}
    \centering
    \includegraphics[width=\linewidth]{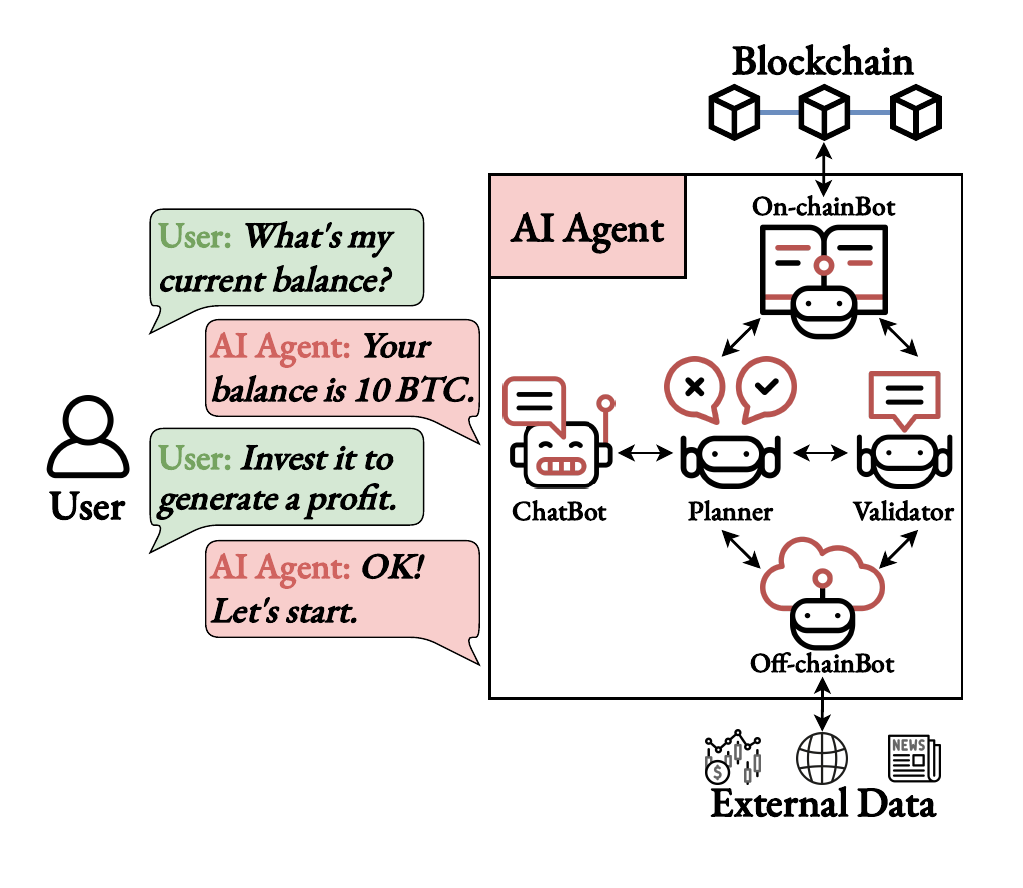}
    \caption{Overview of the interaction between a user and an AI agent for blockchain-related operations.}
    \label{fig:arch}
\end{figure}
The use of Artificial Intelligence (AI)-based agents, powered by Large Language Models (LLMs), has shown remarkable capabilities in task execution and decision-making across a wide range of complex scenarios \cite{10.1145/3580305.3599572}. In recent years, these AI agents have also emerged as valuable tools for interacting with blockchain-based environments. For example, they can be leveraged to detect vulnerabilities within smart contracts \cite{wei2024llm, karanjai2025multi,ma2024combining} or autonomously execute cryptocurrency trades \cite{minarsch2020autonomous, minarsch2021autonomous}. However, such operations often require access to highly sensitive information, including end-users’ private keys, making AI agents a potential attack surface for adversaries. This shift raises significant security and privacy concerns, as integrating AI agents into the Web3 ecosystem may introduce new threat vectors.

Figure \ref{fig:arch} shows a natural language interaction between a user and an AI agent capable of autonomously executing blockchain-related tasks. On the left side, the user asks for their wallet balance and instructs the agent to invest the funds. On the right side, the functional architecture of the agent is depicted, including modules for accessing blockchain data, interpreting user intent, planning tasks, and validating operations to ensure secure and reliable execution. All technical complexity, such as smart contract interaction, API communication, and transaction formatting, is completely hidden from the user, who may have no prior knowledge of blockchain technologies. The agent manages these operations independently and translates user requests into executable actions on the blockchain. It also integrates off-chain information, such as financial news or market trends, to make decisions that are better aligned with current conditions and the user’s goals. This combination of intuitive communication and autonomous execution enables seamless interaction with blockchain systems for both expert and non-expert users.

Several works in the literature \cite{yang2025survey, tran2025multi, wang2025internet} provide general surveys on AI agents but do not adequately address security and privacy concerns. While \cite{wang2025security, deng2025} focus primarily on the security and privacy aspects of these agents, none explicitly consider their application within the blockchain ecosystem. To fill this gap, this paper proposes a novel systematization of knowledge on AI agents in the context of blockchain, with a particular emphasis on security and privacy challenges.

\smallskip
\noindent \textbf{Contributions.} The main contributions of this work are the following:

\begin{itemize}
\item We propose a novel taxonomy of AI agents for blockchain, along with comprehensive reference architectures detailing their components, functionalities, and integration points for effective interaction with blockchain networks.

\item We explore practical applications of AI agents within the blockchain ecosystem, highlighting their roles and benefits, while also thoroughly discussing potential security and privacy vulnerabilities.

\item We systematize practical aspects of AI agents in this domain, including an analysis of the LLMs employed and the datasets used for evaluation.

\item We identify and analyze key open challenges, gaps, and risks associated with deploying AI agents on blockchain, and outline promising research directions.
\end{itemize}

\smallskip
\noindent \textbf{Organization.} The remainder of this paper is structured as follows: Section \ref{sec:back} provides the background on blockchain and LLMs. Section \ref{sec:related} reviews related works in the field. Section \ref{sec:system} introduces a novel taxonomy of AI agents for blockchain and reference architectures. Sections \ref{sec:ai_block} and \ref{sec:ai_smart} discuss how AI agents can streamline operations related to blockchain and smart contracts, respectively. Section \ref{sec:open} identifies open challenges and future research directions. Section \ref{sec:conclusion} concludes the paper.

\section{Background}\label{sec:back}

\subsection{Blockchain}
Blockchain is a decentralized and distributed ledger technology that enables secure and transparent recording of transactions across a network of participants without the need for a trusted central authority \cite{ziyao2019}. Transactions are grouped into blocks that are cryptographically linked using hash functions, ensuring the integrity and immutability of the data. Every participant maintains a copy of the ledger, and the network’s state is updated and agreed upon through decentralized consensus protocols such as Proof of Work (PoW) or Proof of Stake (PoS).
Smart contracts further amplify the potential of blockchain \cite{das2019fastkitten}, which are self-executing programs deployed on the blockchain, whose operation is regulated by the contract terms encoded within them. They have enabled a wide array of applications from DeFi primitives \cite{ozili2022decentralized} to organizational governance through decentralized autonomous organizations (DAOs) \cite{chohan2024decentralized}. Oracles further enrich this ecosystem by providing external data—such as price feeds, weather information, and IoT sensor inputs—enabling smart contracts to respond dynamically to real-world events.

However, despite their transformative potential for many sections, blockchain and smart contract technologies face several challenges that impact their usability and security \cite{8343163, atzei2017survey}. From a user perspective, interacting with blockchain often requires managing cryptographic keys and understanding complex concepts, creating barriers for non-technical users. Moreover, the transparency of blockchain data, while beneficial for auditability, raises privacy concerns. Security vulnerabilities in smart contracts, such as logic errors or improper access controls, have led to significant financial losses in the past and undermine trust in decentralized applications. To mitigate these issues, AI-based agents are emerging as promising tools for their ability to automate complex tasks, such as identifying and even remediating smart-contract vulnerabilities before deployment, and providing intuitive natural-language interfaces to simplify user interactions. However, this integration also expands the threat landscape: AI agents themselves introduce novel attack surfaces and privacy risks, as malicious actors may manipulate autonomous decision-making or exploit agents to extract sensitive information.

\subsection{Large Language Models}
LLMs are advanced neural architectures trained to process and generate human language. Built primarily on transformer-based architectures \cite{vaswani2017attention}, LLMs are pretrained on large-scale corpora encompassing diverse text sources, allowing them to learn complex patterns of syntax, semantics, and factual knowledge. This pretraining enables strong generalization across tasks with minimal or no task-specific supervision. Prominent examples include GPT models \cite{brown2020language,achiam2023gpt}, LLaMA \cite{touvron2023llama}, Mixtral \cite{jiang2024mixtral} and Claude \cite{claude2}. These models achieve state-of-the-art results in various tasks, including text generation, summarization, question answering, translation, and few-shot or zero-shot reasoning. Their performance can be further enhanced through alignment strategies such as supervised fine-tuning and Reinforcement Learning from Human Feedback (RLHF) \cite{ouyang2022training}. A key innovation in extending LLM capabilities is Retrieval-Augmented Generation (RAG) \cite{lewis2020retrieval}, where external knowledge is dynamically retrieved from a corpus and provided as context to the model during inference. This improves factual accuracy, supports access to up-to-date information, and reduces hallucination in tasks that require domain-specific or time-sensitive knowledge. LLMs can also be integrated with tool use, memory, and multi-agent coordination frameworks to support complex reasoning and decision-making pipelines \cite{mialon2023augmented}, further extending their utility across diverse applications.

\subsection{Agent Communication Protocols}
To support seamless interaction, negotiation, and collaboration among AI agents in heterogeneous environments, a range of communication protocols has been introduced. These protocols provide formal mechanisms to standardize interoperability, facilitate integration with external platforms and services, and safeguard data exchange through secure and reliable communication channels.

\subsubsection{Model Context Protocol}
The Model Context Protocol (MCP) \cite{anthropic_mcp} defines a standardized interface for how agents and external tools, services, and data interact. It allows agents to dynamically retrieve, update, and reason over context-specific knowledge, ensuring that interactions remain consistent across heterogeneous environments. MCP supports modularity by decoupling the agent’s reasoning logic from the underlying data sources, thereby enabling flexible integration of domain-specific resources.

\subsubsection{Agent-to-Agent}
Agent-to-Agent (A2A) \cite{a2a_protocol} defines the message exchange mechanisms that govern direct interactions between autonomous agents that need to coordinate and collaborate. It specifies both the structure and semantics of communication, supporting actions such as negotiation, delegation, coordination, and knowledge sharing. Agents are discovered in a structured way through agent cards (i.e., JSON file) shared via trusted registries or endpoints. 

\subsubsection{Agent Network Protocol}
The Agent Network Protocol (ANP) \cite{anp_protocol} facilitates structured communication within multi-agent systems by providing foundational primitives for agent discovery, authentication, and secure data routing. Designed for scalability in large, distributed environments, ANP enables agents to dynamically form coalitions or sub-networks based on task requirements. Each agent is uniquely identified using a Decentralized Identifier (DID) \cite{mazzocca2025survey}, ensuring verifiable identity without centralized control. Communications are protected through end-to-end encryption.

\subsubsection{Agora}
Agora \cite{agora_protocol} is a protocol that facilitates decentralized marketplaces of agents, where participants can publish services, negotiate tasks, and establish trust dynamically. It supports multi-party negotiation and economic coordination through standardized interaction patterns. 

\subsubsection{NANDA}
The Networked Agents and Decentralized AI (NANDA) \cite{nanda_project} protocol is an MIT initiative aiming to provide a structured framework for automated negotiation among agents. NANDA focuses on reaching agreements efficiently by defining standardized phases such as proposal, counter-proposal, acceptance, and commitment. 

\section{Related Work}\label{sec:related}
The integration of blockchain and LLMs holds the potential to address various challenges, ranging from mitigating LLM's security and privacy risks to simplifying user interactions with blockchain systems. Accordingly, the literature includes several surveys that explore this integration from diverse perspectives. Geren et al. \cite{geren2025} focus on analyzing how blockchain has been employed to address LLM-related security concerns.

\begin{figure*}[t!]
\centering
\includegraphics[width=\linewidth]{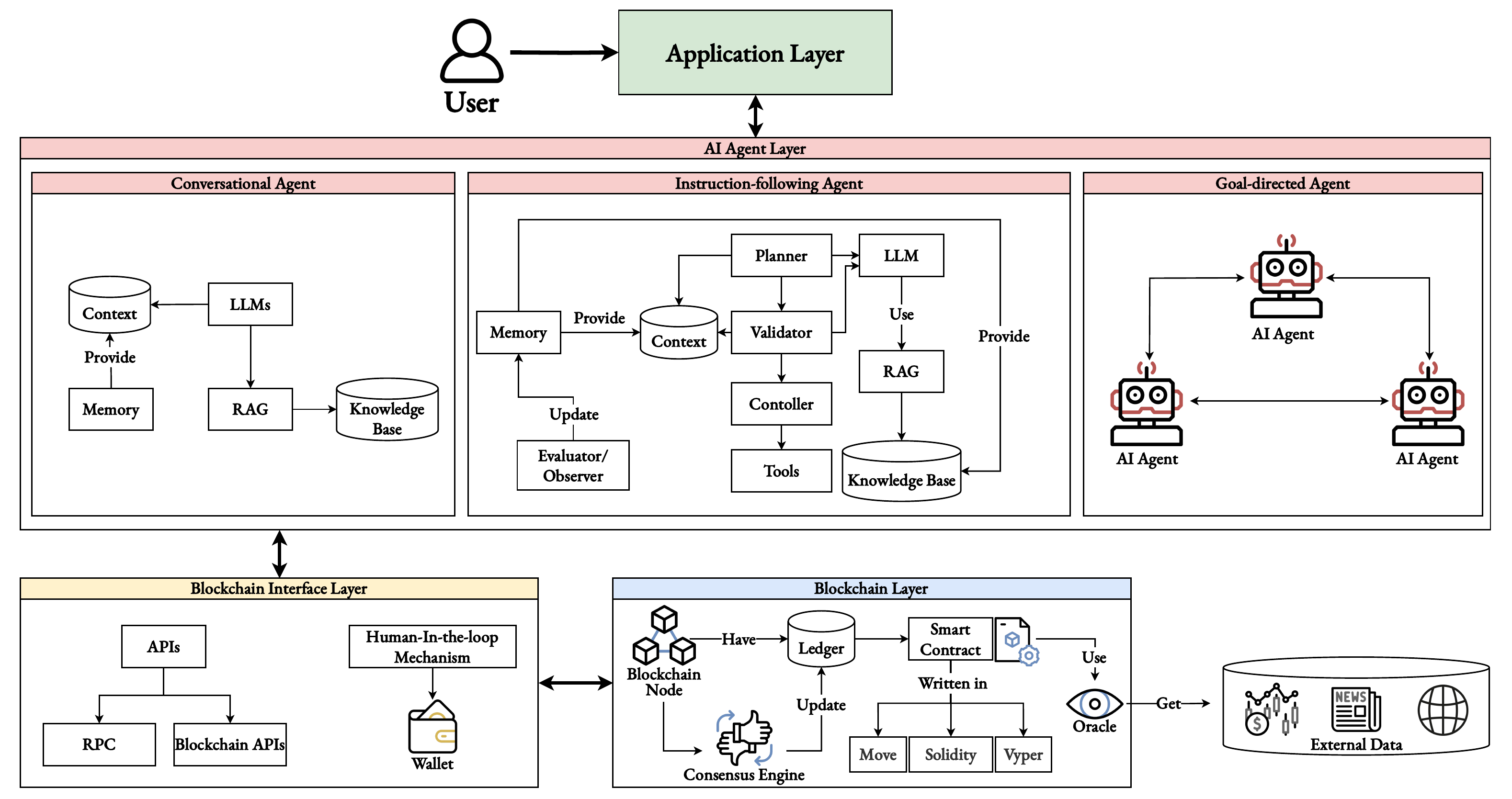}
\caption{Reference architecture of AI4B systems, showing the four-layer structure (Application, AI Agent, Blockchain Interaction, and Blockchain layers) and detailed internal components, including Natural Language Processing, Planning \& Reasoning, Memory Management, Security \& Validation, Tool Integration, APIs, wallet integrations, and blockchain infrastructure.}
\label{fig:detailedarch}
\end{figure*}

Some surveys provide a general overview of AI agents \cite{yang2025survey, tran2025multi, wang2025internet}, which is not tailored for blockchain-based environments nor offers an in-depth discussion on security and privacy issues. While other surveys highlight security and privacy issues of AI agents \cite{wang2025security, deng2025}, they do not explicitly refer to agents employed for interacting with blockchain-based environments. For instance, Wang et al. \cite{wang2025security} only mention blockchain as a valuable tool to audit agents' interactions to enable transparent investigations.  

However, existing works typically consider AI agents in isolation, without anchoring them to specific application domains. We argue that a focused systematization of AI agents within the context of blockchain environments is essential for understanding their role, capabilities, and limitations. To address this gap, our work offers a comprehensive systematization of AI agents for blockchain, with a particular emphasis on security and privacy considerations.

\section{AI4B Taxonomy and Architecture}\label{sec:system}
Given the lack of an established framework for AI agents for Blockchain (AI4B), we first propose a taxonomy that classifies these agents based on their level of autonomy and the nature of user input, ranging from simple questions to complex, high-level goals. Building on this taxonomy, we then present reference architectures for AI agents in blockchain systems, providing a structured foundation to guide their design and implementation.
\subsection{A Taxonomy of AI Agents for Blockchain}
We classify AI agents for blockchain along a spectrum of increasing autonomy, based on how they process user input: as a question to be answered, an instruction to be followed, or a goal to be achieved.
\subsubsection{Conversational Agents} These agents focus on responding to user queries about blockchain data, portfolio status, or protocol parameters. They typically operate in a read-only mode, accessing on-chain and off-chain data sources to provide insights such as wallet balances, recent transactions, token prices, and protocol analytics. The primary function is to enable users to retrieve information through natural language, lowering the barrier to data exploration without requiring technical knowledge of blockchain specifics. Examples include chatbots integrated with DeFi dashboards or wallet applications that allow users to ask questions like “\emph{How much ETH do I currently hold?}”.

\subsubsection{Instruction-following Agents}
These agents go beyond information retrieval by translating explicit user commands into blockchain transactions. They parse natural language instructions such as “Swap 1 ETH for USDC on Uniswap”, transforming them into structured operations that interact with smart contracts. Instruction-following agents often incorporate safety checks, confirmation steps, and parameter validation to prevent costly mistakes. Integration with wallet software enables secure signing and broadcasting of transactions. This class bridges the gap between conversational interfaces and actionable blockchain operations, empowering users to perform tasks without manual interaction with complex user interfaces or command-line tools.

\subsubsection{Goal-directed Agents} The most advanced class of AI agents, goal-directed agents, are designed to autonomously achieve user-defined objectives over multiple steps and time horizons. Instead of simply following a command or answering a query, they plan and execute strategies such as portfolio rebalancing, yield optimization, or risk management across various protocols. These agents continuously monitor blockchain data, market conditions, and user preferences to adapt their actions dynamically. For example, a goal-directed agent might seek to maximize returns by allocating assets among lending platforms while minimizing gas costs and exposure to impermanent loss. Although largely experimental, this category represents a significant step toward fully autonomous DeFi management, where users delegate complex decision-making to intelligent agents.

\subsection{Reference Architectures of AI Agents for Blockchain}

Building on the taxonomy presented in Section~\ref{sec:system}, we introduce a comprehensive reference architecture for AI4B systems, illustrated in Figure~\ref{fig:detailedarch}. This architecture provides a foundational four-layer structure that can be specialized for Conversational, Instruction-following, and Goal-directed agents. We first present the high-level structure and layer responsibilities, then examine the detailed internal components and their interactions.

\subsubsection{Four-Layer Architecture Framework}
As shown in Figure~\ref{fig:detailedarch}, our reference architecture mediates user interactions with on-chain data and smart contracts via four primary layers:

\begin{itemize}
\item \textbf{Application Layer.}  
Serves as the primary interface between users and the AI agent system. This layer encompasses diverse interaction modalities, including Graphical User Interfaces (GUIs), Command-Line Interfaces (CLIs), voice interfaces, and natural language chat systems. The layer's primary responsibility is intent capture—transforming user inputs (whether explicit commands, natural language queries, or high-level goals) into structured, machine-readable requests that the AI Agent Layer can process.
\item \textbf{AI Agent Layer.}  
Functions as the cognitive core of the system, orchestrating all intelligent decision-making processes. This layer performs several critical functions: natural language understanding to parse user intent, context maintenance for conversational continuity, planning and reasoning to decompose complex tasks into executable steps, safety validation to prevent harmful or unintended actions, and tool orchestration to coordinate interactions with external systems. In multi-agent architectures, this layer also manages agent-to-agent communication, task delegation, and collaborative decision-making protocols.
\item \textbf{Blockchain Interaction Layer.}  
Provides the technical infrastructure for all blockchain-related operations, serving as an abstraction layer that shields the AI Agent from the complexities of different blockchain protocols. Core components include RPC/WebSocket clients for node communication, blockchain-specific APIs for protocol interactions, smart contract wrappers and ABIs for contract invocation, data indexers for efficient query processing, wallet integration modules supporting both EOAs and smart wallets, and human-in-the-loop mechanisms for transaction approval and signing.
\item \textbf{Blockchain Layer.}  
Represents the underlying decentralized infrastructure, including blockchain networks, consensus mechanisms, and smart contract execution environments. This layer processes transactions, maintains state consistency, executes smart contract logic, and provides cryptographic guarantees for all operations. It encompasses various blockchain protocols (Ethereum, Polygon, Arbitrum, etc.), different virtual machines (EVM, WASM, etc.), and diverse consensus algorithms (PoS, PoW, etc.).
\end{itemize}

\subsubsection{Detailed Components and Interactions}

Having established the four-layer framework, we now examine the detailed internal components and their interactions within each layer. As defined at the beginning of this section, we identify three AI4B levels, namely Conversational, Instruction-following, and Goal-directed agents, all of which conform to the four-layer structure described above. The Application, Blockchain Interaction, and Blockchain Layers remain unchanged across these classes. While Conversational agents typically perform only read-only operations, and Instruction-following and Goal-directed agents must also construct and submit transactions to achieve their objectives, all three archetypes share the same sub-module pattern. The key distinctions arise within the AI Agent Layer itself.

Cheng et al. \cite{cheng2024exploring} organize the basic framework of LLM-based agent framework as five components: Planning, Memory, Rethinking, Environments, and Action. AI agents plan, rethink their plans, use memory in these processes as context or knowledge base, and interact with environments by observation or executing actions using tools. This basic framework aligns with our reference framework. Figure~\ref{fig:detailedarch} captures these variations in our detailed reference architecture for the AI Agent Layer, drawing on several seminal designs: the modular, Web3-native OS of Eliza \cite{walters2025eliza}, the multi-agent chatbot framework of Nguyen et al. \cite{nguyen2024multi}, the decentralized autonomous collaboration model of DeCoAgent \cite{jin2024decoagent}, Supply Chain Traceability systems \cite{santos2025using}, and others including Crypto Portfolio Management agents~\cite{luo2025llm}. 

The Application Layer serves as an interface for user inputs, typically through app UX \cite{walters2025eliza} and/or natural language interfaces \cite{benzinho2024llm, nguyen2024multi}. It wraps the user intent or goal as a query into a structured request for the AI Agent.

The AI Agent Layer processes this request through its internal components. Simple conversational agents have a straightforward structure composed of an LLM model with memory that provides a knowledge base and context, while using RAG or other techniques to improve the response \cite{santos2025using}. Instruction-following agents \cite{walters2025eliza, nguyen2024multi} perform more complex processing, often necessitating a more sophisticated architecture. As Cheng et al. \cite{cheng2024exploring} organize, they have a common structure composed of a Planner, which uses LLM models such as GPT-4o and Claude 3.5 Sonnet for understanding user natural language queries and developing plans. 

\smallskip
\noindent \textbf{Enhanced Context.}
Some AI4B systems use techniques such as RAG or more advanced memory-based context input \cite{walters2025eliza} as context to improve their internal processes. Then, the Validator component evaluates plans and refines them. For instance, ElizaOS~\cite{walters2025eliza} implements Trust Score Managers and Evaluators that assess proposed blockchain operations by calculating risk scores based on token performance and market conditions, automatically blocking transactions below configurable trust thresholds to prevent harmful operations. This layer can also function as a privacy module to ensure that the agent will not harm the user. To execute the validated plans, the agent needs to identify the appropriate tools and determine which to call—here, the Tool Mapper/Controller module operates. It enables the agent to select the proper tools. The Evaluation/Observation Module receives feedback from the environment, typically the blockchain in our context, and updates the memory while providing feedback to the agent to achieve its goals. Lastly, Goal-based agents, such as Crypto Portfolio Management agents \cite{luo2025llm}, work by having these agents interact and delegate to each other to achieve the goals.

The Blockchain Interaction Layer facilitates blockchain interactions through tools, enabling the AI agent to alter or retrieve chain state, as seen when executing smart contracts. For instance, Nguyen et al. \cite{nguyen2024multi} implement wallet integration through MetaMask connectivity, maintaining familiar wallet-based transaction approval flows as the interaction layer. Other frameworks use blockchain APIs that embed chain interactions within themselves. ElizaOS \cite{walters2025eliza} connects to various blockchain networks through RPC endpoints and specialized APIs. Additionally, Santos et al. \cite{santos2025using} implement an API Gateway pattern that fetches processed data from blockchain-connected services rather than making direct blockchain calls.

The final infrastructure layer, the Blockchain Layer, places transactions into blocks, achieves consensus, and enables blockchain nodes to execute smart contracts through transactions, thereby changing states. Index nodes return chain state information, such as smart contract states via RPC nodes, providing feedback to the upper layers.

It is worth noting that security and privacy modules exist across multiple layers. For instance, Nguyen et al. \cite{nguyen2024multi} use a human-in-the-loop mechanism by wallet integration. Others \cite{walters2025eliza, luo2025llm} use Validator and Observer components at the AI agent layer for this purpose. Specifically, Walters et al. \cite{walters2025eliza} implement Trust Score Managers that calculate risk scores before financial operations as their Validator component. Similarly, Luo et al. \cite{luo2025llm} employ a dual-validation approach incorporating confidence scoring based on token probabilities from their Observation/Evaluation Module, plus validation through intra-team agent collaboration, where agents vote on decisions based on their respective confidence levels.

\noindent \textbf{Non-Normative Example.} 
The AI4B system provided by Nguyen et al.~\cite{nguyen2024multi} is an example of an Instruction-following agent (see Table \ref{tab:ai_blockchain_works}). Specifically, it enables users to interact with a blockchain intuitively through chat, from reading blockchain data to executing transactions. In this system, the user provides prompts through the chat UX (the Application Layer), which the AI Agent Layer receives. As Memory in the Agent Layer, the user's conversation history, examples of query refinements (pairs of user query inputs and their refined outputs), and tool descriptions (blockchain API descriptions and input parameter specifications) are stored as embedded vectors. Upon user prompt input, the prompt refining agent refines a user prompt using conversation history and examples of query refinements. It passes the refined prompt to the initial tool filtering component, which uses the tool descriptions and passes the likely tool list to the planning agent. The agent (Planner in our AI4B reference architecture) receives the refined prompt and likely tool list as context and then makes a plan of actions including what tools to call~\footnote{The tool descriptions and tool calling mechanisms in this implementation could be replaced by modern standardized protocols such as MCP or A2A, which were not as widely adopted at the time of this system's development.}. The action plan and tool calling are validated by the LLM-based validation agent (Validator in our reference architecture), and the same agent will call validated tools (APIs). In this implementation, the Validator also serves as the Controller component. The tools invoke blockchain APIs, including data-retrieval APIs and transaction-execution APIs, which correspond to the Blockchain Interaction Layer. The tool execution results are received by the LLM-based evaluation component (Evaluator in our reference architecture). The Evaluator then checks whether the user's intention from the prompt has been satisfied based on the current state and generates a response. Finally, the user receives the response through the chat.

\begin{table*}[t!]
\centering
\caption{Representative AI-Agent Solutions for Interacting with Blockchain}
\begin{adjustbox}{width=\textwidth}
\begin{tabular}{llllp{4.5cm}c}
\toprule
\textbf{Type} & \textbf{Work} & \textbf{Year} & \textbf{Use Case} & \textbf{LLM} & \textbf{Open Source} \\
\midrule
\multirow{3}{*}{Conversational} 
& Toyoda~\textit{et al.}~\cite{toyoda2024blockchain} & 2024 & Analyzing and Interacting with Blockchain Data & N/A & \xmark \\
& Benzinho~\textit{et al.}~\cite{benzinho2024llm} & 2024 & Supply Chain Traceability & Mixtral 8x7B, Mixtral 7B, LLaMA 3 8B \& Gemma 2B & \xmark \\
& Santos~\textit{et al.}~\cite{santos2025using} & 2025 & Supply Chain Traceability & N/A & \xmark \\
\midrule
\multirow{3}{*}{Instruction-following} 
& Nguyen~\textit{et al.}~\cite{nguyen2024multi} & 2024 & Analyzing and Interacting with Blockchain Data & GPT-4o & \codeicon{https://github.com/Duc-NSH/blockchain-chatbot} \\
& Walters~\textit{et al.}~\cite{walters2025eliza} & 2025 & Analyzing and Interacting with Blockchain Data & N/A & \codeicon{https://github.com/elizaOS/eliza} \\
& Ao~\textit{et al.}~\cite{ao2025agentdao} & 2025 & DAO Governance \& Coordination & GPT-4o & \xmark \\
\midrule
\multirow{4}{*}{Goal-Directed} 
& Minarsch~\textit{et al.}~\cite{minarsch2020autonomous} & 2020 & Portfolio Management \& DeFi Trading & N/A & \codeicon{https://github.com/fetchai/agents-aea} \\
& Minarsch~\textit{et al.}~\cite{minarsch2021autonomous} & 2021 & Portfolio Management \& DeFi Trading & N/A & \codeicon{https://github.com/fetchai/agents-aea} \\
& Li~\textit{et al.}~\cite{li2024reflective} & 2024 & Portfolio Management \& DeFi Trading & GPT-3 \& GPT-4o & \codeicon{https://github.com/Xtra-Computing/CryptoTrade} \\
& Luo~\textit{et al.}~\cite{luo2025llm} & 2025 & Portfolio Management \& DeFi Trading & GPT-4o & \xmark \\
\bottomrule
\end{tabular}
\end{adjustbox}
\label{tab:ai_blockchain_works}
\end{table*}

\section{AI Agents for Interacting with Blockchain}\label{sec:ai_block}
The proliferation of blockchain-based systems has created powerful new paradigms for decentralized applications (dApps), finance (DeFi), and governance (DAOs). However, the technical complexity inherent in these systems remains a formidable barrier to mainstream adoption. Users are often required to understand cryptographic principles, manage private keys, interpret complex smart contract logic, and navigate unintuitive interfaces. This usability gap has driven a new field of research and development focused on employing AI agents to serve as intelligent intermediaries, abstracting away complexity and making blockchain technology more accessible, secure, and efficient. In this section, we systematize the emerging landscape of AI agents designed for blockchain interaction. 
Table \ref{tab:ai_blockchain_works} provides an overview of representative AI-agent solutions designed to interact with blockchain systems, categorized by agent type, application domain, publication year, the LLM used, and availability of open-source implementations.

\subsection{Applications and Use Cases}

\subsubsection{Analyzing and Interacting with Blockchain Data}
AI agents act as intelligent interfaces to explore, interpret, analyze, and interact with on-chain data, lowering the technical barrier for users unfamiliar with blockchain query languages or analytics tools. Using natural language processing and RAG, these agents let users ask open-ended questions about blockchain activity and receive clear, structured answers. Beyond simple queries, they can detect anomalies such as fraudulent transactions, supporting auditing, forensic analysis, protocol monitoring, and user behavior tracking. For example, agents can summarize wallet histories, flag suspicious token movements, generate real-time alerts, and sometimes assist with executing transactions.
Recent work by Toyoda~\textit{et al.}~\cite{toyoda2024blockchain} offers a systematic exploration of how LLM integration addresses key challenges in blockchain analytics, including data scarcity, protocol heterogeneity, and limited explanation of model outputs. They propose a framework that enhances anomaly detection by combining prompt engineering, Chain-of-Thought reasoning, and modular “predictor” and “enhancer” design patterns. The authors demonstrate that LLMs can provide improved generalizability across chains, better interpretability compared to rule-based systems, and natural language explanations of smart contract behavior. These agents can surface signs of fraud and identify protocol vulnerabilities far more flexibly than traditional analytics engines.
The system presented in \cite{nguyen2024multi} introduces a modular multi-agent architecture that streamlines both conversational access to blockchain data and transaction execution. It includes agents for query refinement and API selection based on embedding similarity. By supporting both data exploration and smart contract interaction, the system reduces the complexity of blockchain APIs for non-technical users.
ElizaOS \cite{walters2025eliza} introduces a full-stack operating system for autonomous AI agents on blockchain, focusing on persistent memory, multi-modal reasoning, and decentralized execution. It provides native support for long-running agents capable of observing on-chain events, invoking smart contract methods, and adapting behavior over time. ElizaOS focuses on modular agent design and flexible integration tools, allowing developers to define custom workflows that span from natural language queries to on-chain actuation.

\subsubsection{Supply Chain Traceability}  A particularly valuable application of AI agents in blockchain is supply chain traceability. These agents support full traceability across diverse supply chains, including sectors such as food, pharmaceuticals, and electronics. These agents enable users to query and interpret on-chain records related to production, transport, and certification, presenting complex provenance data in natural language. Moreover, these agents can automate actions such as recording new supply chain events or validating certifications on-chain. This enhances transparency, supports regulatory compliance, and facilitates the detection of fraud or inefficiencies. For example, the works in \cite{benzinho2024llm,santos2025using} present a farm-to-fork traceability agent powered by RAG, enabling consumers to seamlessly query detailed provenance, processing, and logistical information stored on an immutable blockchain ledger.

\subsubsection{Portfolio Management \& DeFi Trading}
AI agents in finance play a central role in blockchain applications, particularly in asset management, DeFi, and autonomous economic activities. By abstracting technical complexity, such agents lower the barrier to entry for non-expert users and enable more informed, data-driven strategies in decentralized financial ecosystems. Portfolio management agents assist users in monitoring asset holdings, rebalancing portfolios, and optimizing risk-return tradeoffs through machine learning models that analyze both on-chain and off-chain data \cite{luo2025llm}. DeFi and trading agents automate complex interactions with multiple protocols, such as yield farming, liquidity provision, and arbitrage opportunities, often employing reinforcement learning or multi-agent coordination to maximize returns. The authors in \cite{li2024reflective} introduce CryptoTrade, an LLM-based trading agent for cryptocurrency markets that combines the analysis of on-chain data and off-chain signals, such as financial news. Unlike prior work focused on stock trading, CryptoTrade leverages blockchain transparency and incorporates a reflective mechanism to refine decisions based on past performance. A more advanced class of agents in this domain is represented by Autonomous Economic Agents (AEAs) \cite{minarsch2020autonomous, minarsch2021autonomous}, which are capable of making fully independent economic decisions and executing transactions on blockchain platforms. These agents continuously monitor markets, negotiate with other agents, and dynamically adapt strategies without human intervention. Frameworks like Fetch.ai \cite{fetchFetchaiBuild} and Olas \cite{olasOlasCoown} provide the infrastructure to develop such agents, which are expected to play a foundational role in enabling decentralized machine-to-machine economies.

\subsubsection{DAO Governance \& Coordination}
DAOs rely on collective decision-making processes that can be complex and time-consuming for participants. AI agents are increasingly used to facilitate governance and coordination within DAOs by automating tasks such as proposal summarization, discussion moderation, voting assistance, and outcome prediction. These agents help reduce information overload by extracting key insights from large volumes of textual proposals and community discussions, making it easier for members to engage meaningfully. For example, AgentDAO \cite{ao2025agentdao} uses a multi-agent LLM system augmented with a label-centric retrieval algorithm and a domain-specific proposal language to translate natural language inputs into executable DAO proposal payloads.

\subsection{Threats and Risks}

\subsubsection{Erroneous Behavior and Financial Losses}
Misinterpretation of user intent by AI agents can result in incorrect or unintended transactions. As pointed out in \cite{izadi2024error}, even minor ambiguity can cause costly errors when moving digital assets. The irreversible nature of blockchain operations means such mistakes may have lasting financial consequences.

\subsubsection{Security Vulnerabilities and Exploits}
AI Agents that manage wallet access and interact with DeFi protocols create new attack surfaces. Unauthenticated or overly privileged agent layers can expose users to wallet draining or unauthorized access \cite{cointelegraphAgentsComing}. Potential breaches could result from attack vectors such as prompt injection \cite{liu2023prompt,greshake2023not}.

\subsubsection{Context Manipulation and Memory Attacks}
Patlan~\textit{et al.}~\cite{patlan2025real} demonstrate that AI agents relying on internal memory are vulnerable to “fake memory” or context manipulation attacks, where adversaries inject malicious memory to deceive the agent into harmful transaction flows. This is especially dangerous for goal-directed agents that maintain a persistent dialogue context.

\subsubsection{Privacy Leakage}
To function effectively, AI agents often require access to sensitive on-chain data, such as wallet addresses, transaction histories, and user behavior across protocols. AI agents may inadvertently expose this information through logs, conversational transcripts, or model outputs \cite{kim2023propile}. In \cite{zharmagambetov2025agentdam}, the authors show that LLM-powered agents routinely access more private information than necessary.

\subsubsection{Autonomy-Induced Market Risks}
Autonomous agents executing multiple trades can unintentionally increase market volatility or cause systemic risks \cite{min2022systemic}. AI-driven trading could amplify shocks, collude across platforms, or reinforce herd behavior. Interconnected or correlated algorithms may exacerbate flash-crash events \cite{canonico2019flash}.

\subsubsection{Over‑Reliance and Loss of Human Oversight}
Excessive trust in AI agents can lead users to abdicate responsibility for critical financial decisions. Widespread agentic behavior may degrade vigilance, allowing malicious or flawed actions to go unchecked. 

 \begin{tcolorbox}
    \noindent  \textbf{T-1. Takeaways on AI Agents for Blockchain.} AI agents represent a promising approach to bridging the usability gap in blockchain systems by abstracting technical complexity and enabling more intuitive interaction. Their impact is already evident across key domains such as supply chain tracking, DeFi, DAO governance, and autonomous economic coordination. However, as AI agents gain autonomy and deeper integration with blockchain, they introduce novel risks, including erroneous actions with financial consequences, new attack vectors targeting agent security, vulnerabilities to context manipulation, privacy leakage, market destabilization, and reduced human oversight.
    \end{tcolorbox}

\begin{table*}[ht]
\centering
\caption{Representative AI-Agent Solutions for Smart Contract Development}
\begin{tabular}{lclp{4.5cm}p{5cm}c}
\toprule
\textbf{Work} & \textbf{Year} & \textbf{Use Case} & \textbf{LLM} & \textbf{Dataset} & \textbf{Open Source} \\
\midrule
Qasse~\textit{et al.}~\cite{icontractbot} & 2021 & Development & N/A & N/A& \xmark \\
Qasse~\textit{et al.}~\cite{chat2code} & 2023 & Development & N/A & N/A & \codeicon{https://zenodo.org/record/7391855} \\

Petrović~\textit{et al.}~\cite{petrovic2023model} & 2023 & Development & GPT-3.5 & N/A & \codeicon{https://github.com/ialazzon/chatgptSC} \\
Wei~\textit{et al.}~\cite{wei2024llm} & 2024 & Auditing & GPT-3.5 \& GPT-4o & Custom labeled dataset (110 contracts across 11 vulnerability classes) and Real-world dataset (102 projects, 6,454 contracts)
derived from Code4rena \cite{zhang2023demystifying} & \codeicon{https://github.com/LLMAudit/LLMSmartAuditTool} \\
Ma~\textit{et al.}~\cite{ma2024combining} & 2024 & Auditing & Gemma 7B & Custom balanced dataset of 3,544 samples from 263 audit reports, including 1,734 vulnerable functions with reasoning and 1,810 benign functions
& \codeicon{https://drive.google.com/drive/folders/1uSXaY7vOvcwQIwXs5JwD9C2hxK9bFMsZ} \\
Luo~\textit{et al.}~\cite{FELLMVP} & 2024 & Auditing & Gemma 7B & Dataset from \cite{liu2023rethinking} comprising 15,637 labeled samples across 8 vulnerability types, including 820 positive and 14,817 negative examples & \codeicon{https://drive.google.com/drive/folders/1uSXaY7vOvcwQIwXs5JwD9C2hxK9bFMsZ} \\
Jiang~\textit{et al.}~\cite{gas-wasting} & 2024 & Auditing & GPT-4 & Dataset of 6,614 Solidity contracts and 26 Vyper contracts from Etherscan \cite{etherscanEthereumETH}(August 2023), refined through multi-step filtering to 311 high-complexity, non-redundant contracts for gas-inefficiency analysis & \codeicon{https://github.com/jinan789/gas_patterns} \\
Luo~\textit{et al.}~\cite{luo2025guiding} & 2025 & Development & GPT-4o, Gemini 1.5-Flash, Qwen-plus, Llama 3.1 8B/405B, \& Qwen2.5 7B & Custom open-source fine-tuning dataset (30k samples) constructed from smart contracts collected via Etherscan \cite{etherscanEthereumETH}, paired with GPT-4o-generated user requirements. Designed to support diverse smart contract generation tasks in a dialog-based format. & \xmark \\

Karanjai~\textit{et al.}~\cite{karanjai2025multi} & 2025 & Auditing & Gemma 2 9B \& CodeGemma & Collection of 60 intentionally vulnerable Solidity contracts from Trail of Bits' "Not-So-Smart Contracts" repository \cite{githubGitHubCryticnotsosmartcontracts} and 92 real-world Move projects (652 modules) sourced from Aptos, Sui, and Starcoin ecosystems & \xmark \\

Li~\textit{et al.}~\cite{li2025scalm} & 2025 & Auditing & GPT-4, GPT-4o, Claude 3.5 Sonnet, Gemini 1.5 Pro, \& Llama 3.1 70B & DAppSCAN dataset \cite{zheng2024dappscan} (39,904 Solidity files with 1,618 SWC weaknesses from 682 projects) and SmartBugs dataset \cite{durieux2020empirical} (1,894 contracts with 5 SWC categories) 
 &\codeicon{https://figshare.com/s/5cc3639706e4ecd16724} \\

\bottomrule
\end{tabular}
\label{tab:ai_blockchain_sc}
\end{table*}

\section{AI Agents for Smart Contract Development}\label{sec:ai_smart}
Smart contract is a key technology in Web3 as they enable the deployment of self-executing programs directly on the blockchain. They are supported by multiple blockchain platforms and can be implemented in various programming languages, such as Solidity or Rust. However, creating smart contracts from scratch can be challenging, especially for developers with limited experience. In this context, AI agents can play a fundamental role by supporting developers throughout the smart contract lifecycle. Table \ref{tab:ai_blockchain_sc} summarizes representative AI-agent solutions aimed at supporting smart contract development and auditing. The solutions are categorized by publication year, use case, LLM employed, datasets utilized, and availability of open-source resources.

\subsection{Application and Use Cases}

\subsubsection{Development}
Smart contracts were originally proposed by Nick Szabo in the mid‑1990s as digital transaction protocols designed to enforce the terms of a contract automatically \cite{szabo1996smart}. However, lawyers and other legal professionals responsible for drafting traditional agreements often lack the technical expertise required to implement an equivalent digital version as a smart contract. In this context, AI agents can play a crucial role by translating high‑level requirements and policies into formal smart contract specifications across different programming languages \cite{icontractbot,petrovic2023model}. Chatbots, such as Chat2Code \cite{chat2code}, further facilitate this process by providing an interactive environment where domain experts can collaborate with AI agents to refine and implement the desired smart contract iteratively.

To better align smart contract generation with user requirements, formal methods have been proposed as a solution. For example, Luo~\textit{et al.}~\cite{luo2025guiding} integrate Finite State Machines (FSMs) as a formal modeling technique to precisely capture high‑level requirements and guide the automated generation process, thereby reducing errors, increasing reliability, and ensuring closer conformance to the intended design.


\subsubsection{Auditing}
While transparency and immutability are two defining strengths of smart contracts, they also represent inherent limitations. Specifically, if a smart contract contains a vulnerability, the only option to address it is to deploy a new, updated version. Moreover, because smart contracts are fully transparent and publicly accessible, adversaries can analyze their code to identify and take advantage of weaknesses. There are many incidents, such as the DAO Hack, where vulnerabilities in smart contracts have been successfully exploited \cite{liya}.

Given the smart contract's central role in the Web3 ecosystem, ensuring they do not contain hidden vulnerabilities is critical. Intuitively, AI agents offer promising capabilities to assist in this task by automatically detecting potential flaws and enhancing contract security before deployment. 

Multi‑agent applications have demonstrated superior effectiveness compared to single‑agent systems across a range of domains \cite{wu2023autogen}. Building on this insight, multi‑agent approaches have also been adopted for detecting vulnerabilities in smart contracts. For example, Wei~\textit{et al.}~\cite{wei2024llm} introduce LLM‑SmartAudit, a multi‑agent conversational system that employs a team of specialized agents to identify security flaws within smart contract code. Similarly, Karanjai~\textit{et al.}~\cite{karanjai2025multi} present Smartify, a multi‑agent framework that leverages a set of fine‑tuned LLM‑based agents, each focusing on distinct vulnerability detection tasks, to enable highly automated and accurate smart contract analysis. Ma~\textit{et al.}~\cite{ma2024combining} also highlight the benefits of fine‑tuning, proposing a two‑stage fine‑tuning strategy. In this approach, a Detector model is first fine‑tuned to focus exclusively on identifying vulnerabilities, while a Reasoner model is subsequently fine‑tuned to explain the root causes behind the detected issues. Luo~\textit{et al.}~\cite{FELLMVP} propose FELLMVP, a novel framework that combines ensemble learning with LLMs to improve vulnerability detection in Solidity smart contracts. Notably, it introduces a novel text-representation, which captures both the internal structure of a contract and its external call relationships, allowing LLMs to better understand the semantics, internal behavior, and external interactions of smart contract code.

Additionally, AI agents can be used to detect bad practices that, while not directly causing security incidents, can still raise the risk of future issues. For example, SCALM \cite{li2025scalm} combines step-back prompting and RAG to enable this type of detection. It performs static analysis across large codebases to identify and extract code blocks associated with potential bad practices, then vectorizes and stores them in a searchable knowledge base. By leveraging RAG and step‑back prompting, SCALM can abstract higher‑level concepts and coding principles, making it possible to pinpoint problematic patterns more effectively. 


\subsubsection{Deployment}
Deploying a smart contract is challenging because it is an irreversible process that must be executed correctly the first time, and any bug or design error becomes permanently embedded. It also involves gas costs, which vary with network congestion and contract size, leading to economic and temporal constraints that must be carefully managed. Ensuring compatibility with the target network’s virtual machine and protocol standards is also critical; any discrepancies can result in deployment failures or limit interoperability across blockchains.

AI agents can serve as valuable tools in mitigating these challenges. By analyzing contract logic and usage patterns, they can suggest optimizations that enhance gas efficiency, thereby reducing deployment costs \cite{gas-wasting}. Moreover, these systems can evaluate platform-specific constraints, identify versioning issues or misconfigurations, and recommend necessary adjustments to ensure successful deployment and broader platform compatibility \cite{8847638}.


\subsection{Threats and Risks}

\subsubsection{Domain‑Specific Language Bias}
LLMs such as GPT are primarily trained on publicly available data, which predominantly includes code written in popular programming languages like Python, JavaScript, and C++ \cite{li2024quantifying}. In contrast, smart contract languages such as Solidity, Vyper, and Move have smaller developer communities and fewer available training examples, making AI‑generated code more prone to syntax and logical errors. 

\subsubsection{Model Dependence and Adaptability}
The effectiveness of auditing systems is closely tied to the capabilities and limitations of the underlying LLMs. As discussed earlier, these systems inherit the knowledge and biases of their training data, making their performance highly dependent on the quality, completeness, and relevance of that data. As a result, an auditing tool may fail to detect certain vulnerabilities if they were not adequately represented during training, or it may misclassify benign patterns as threats if its internal decision boundaries are too strict.

Moreover, as new vulnerability types and attack techniques evolve, the performance of auditing systems may degrade unless their underlying models are retrained or fine‑tuned with updated information. This strong dependence on a specific model introduces a risk of over‑reliance, creating a single point of failure and limiting the system’s ability to adapt to the dynamic threat landscape of smart contract security.

\subsubsection{Hallucinations}
In multi‑agent smart contract auditing, the output of one agent may be tainted by hallucinations or influenced by adversarial inputs, and subsequently consumed as input by other agents. This propagation of inaccurate or manipulated information can create a cascading effect, where misclassifications or fabricated findings are compounded across the collaboration chain. Over time, such error accumulation can degrade the precision of vulnerability detection, obscure critical security flaws, and ultimately undermine the reliability and trustworthiness of the entire multi‑agent auditing process.

\subsubsection{Sensitive Information Exposure}
Deploying smart contracts often involves handling or processing sensitive information, such as user data, and accessing a private wallet. AI agents that support deployment and auditing may require access to this information, introducing the risk of data leakage or unauthorized access. This is especially critical when relying on third‑party or cloud‑hosted AI services, where sensitive data might be exposed to external environments. These concerns are commonly overlooked in the literature. 
\subsection{Benchmark Datasets}
While several datasets are commonly used in smart contract security research, their quality, limitations, and reproducibility vary significantly. Small annotated datasets, such as Not-So-Smart Contracts \cite{githubGitHubCryticnotsosmartcontracts} or SmartBugs Curated \cite{durieux2020empirical,githubGitHubSmartbugssmartbugscurated}, provide clear labels across well-defined vulnerability categories, enabling controlled benchmarking. However, these datasets often lack the complexity and scale of real-world contracts, which may reduce their relevance for practical use cases. Large-scale crawled datasets, including DAppSCAN \cite{zheng2024dappscan},  SmartBugs Wild \cite{durieux2020empirical,githubGitHubSmartbugssmartbugswild}, and collections from Etherscan \cite{etherscanEthereumETH} and Code4rena contest \cite{zhang2023demystifying}, reflect real-world coding practices and diverse vulnerabilities. These datasets are valuable for evaluating models under realistic conditions, but they often contain duplicates, skewed distributions, or noise, which may bias evaluation outcomes. Furthermore, most datasets primarily focus on Solidity, with Vyper \cite{gas-wasting} and Move contracts \cite{karanjai2025multi} being represented only sparsely. Severity annotations are rare, limiting risk-aware analyses \cite{wei2024llm}. In terms of reproducibility, most datasets 
are publicly available, which facilitates independent validation. Nevertheless, preprocessing steps such as deduplication, compiler version filtering, or partitioning are not always standardized or thoroughly documented, which can complicate replication. Audit-based datasets may impose licensing restrictions or grant only partial access to the audit reports \cite{ma2024combining}.

 \begin{tcolorbox}
    \noindent  \textbf{T-2. Takeaways on AI Agents for Smart Contracts.} Most AI agents focus on smart contract vulnerability detection, often tailored to specific programming languages. However, the absence of a common reference benchmark raises concerns about their generalizability, as results may be influenced by the specific training and fine‑tuning of the underlying LLMs. Moreover, while deploying smart contracts is challenging due to the diverse configurations and environments across blockchain platforms, no AI agents have yet addressed this critical aspect.
    \end{tcolorbox}
\section{Open Challenges \& Research Directions}\label{sec:open}

This section identifies open challenges and future research directions that hold the promise to enhance the effectiveness of AI agents for blockchain, enhancing their widespread adoption.

\subsection{Limitations of Current Datasets}
AI‑based agents for smart contract development and auditing would benefit greatly from common, high‑quality, and representative benchmark datasets, which currently remain a critical bottleneck \cite{daspe_benchmark}. At present, results are often evaluated on privately curated or highly specific datasets, making it challenging to assess their generalizability or reliably compare different approaches. The establishment of openly available, community‑endorsed benchmarking platforms, akin to those used in other areas of AI research, would enable fair, reproducible comparisons and foster incremental advances across the field. This need is particularly pressing for low-resource smart contract languages such as Vyper and Rust, which are gaining traction due to their security-focused design and suitability for formal verification. Despite their growing relevance, these languages suffer from a lack of annotated datasets and tooling support compared to more established languages like Solidity. As a result, AI agents trained on Solidity-centric corpora may struggle to generalize across languages, limiting their utility in diverse blockchain ecosystems.

\subsection{Privacy-Preserving AI Agents}
Many of the tasks that AI agents can streamline, from contract deployment to transaction management, require access to highly sensitive information, such as private keys or credentials stored in a user’s wallet. To enable this securely, it is essential to define new protocols and access methods that balance automation with trust and privacy guarantees. Protocols such as MCP or Agent‑to‑Agent A2A can serve as foundational building blocks, allowing AI agents to interact with wallets and sensitive data in a secure, auditable, and user‑controlled manner \cite{wang2025security}.

\subsection{Accountability and Auditability of AI Agents}
AI agents operating in blockchain environments require robust accountability mechanisms, particularly given the irreversible nature of blockchain transactions and the financial implications of DeFi operations. While validation modules~\cite{walters2025eliza} and human-in-the-loop mechanisms~\cite{nguyen2024multi} provide safety guards during execution, there is a critical need for comprehensive audit trails that enable post-hoc analysis of agent decisions and actions. Current systems lack standardized frameworks for recording decision rationales, tracking the chain of reasoning that led to specific actions, and maintaining immutable logs of agent behavior that can be examined after incidents occur~\cite{south2025authenticated}. Future research should focus on developing blockchain-native audit systems that capture not only the actions agents took but also the reasons behind them, including contextual information, risk assessments, and decision criteria used. Such accountability frameworks become especially crucial as agents gain autonomy in managing significant financial assets, where understanding the root cause of failures or suboptimal decisions is essential for user trust and regulatory compliance.

\subsection{User Experience and Human-AI Collaboration}
The design of effective user experiences for AI agents in blockchain systems presents unique challenges in balancing automation with user control and consent. Unlike traditional applications, blockchain interactions involve irreversible financial transactions, making the timing and granularity of human involvement critical design decisions. Currently, established guidelines for when and how to obtain user permissions, how to present complex blockchain operations in understandable terms, and how to maintain user agency while enabling seamless automation do not exist. Key research directions include developing standardized interaction patterns for consent management, designing intuitive interfaces that help users understand the implications of delegating decisions to AI agents, and establishing best practices for progressive disclosure of agent capabilities. Furthermore, research is needed on adaptive interaction models that can adjust the level of human involvement based on transaction risk, user expertise, and contextual factors. The goal is to create AI agents that enhance user capabilities without sacrificing transparency, control, or understanding of the underlying blockchain operations.

\subsection{Multi-Chain Agents}
As multi‑chain ecosystems evolve, the effectiveness of AI agents will increasingly hinge on their ability to operate seamlessly across diverse protocols, virtual machines, and execution environments. The future lies in multi‑chain AI agents that can understand both the nuances and commonalities between platforms, from Ethereum and its EVM‑compatible chains to ecosystems like Cosmos \cite{cosmos} with its Inter‑Blockchain Communication (IBC) protocol \cite{ibc}. Such agents would enable developers and auditors to work efficiently across fragmented Web3 landscapes, making it possible to design, verify, and maintain smart contracts and decentralized applications that span multiple chains.
\section{Conclusion}\label{sec:conclusion}
AI agents are transforming numerous applications, bringing a spectrum of capabilities and inherent risks. While general AI systems have been widely explored, this paper addresses a critical gap by presenting the first comprehensive systematization of AI agents within blockchain environments. Our analysis highlights key challenges, limitations, and opportunities of this integration, laying the groundwork for future research and guiding the secure, trustworthy, and responsible deployment of AI in decentralized ecosystems.

\section*{Acknowledgment}
This work was partially supported by the SERICS (PE00000014) project under the MUR National Recovery and Resilience Plan program funded by the European Union - NextGenerationEU.

\bibliographystyle{IEEEtran}
\bibliography{main} 

\end{document}